\documentclass[letterpaper,draft]{appolb}
\usepackage{epsfig}

\begin{document}
\def\SS{{\cal S}}
\def\R{{I\kern-.30em{R}}}
\def\Z{{Z\kern-.5em{Z}}}
\def\P{{I\kern-.30em{P}}}
\def\N{{I\kern-.22em{N}}}
\def\E{{I\kern-.25em{E}}}

\def\a{\alpha}
\def\b{\beta}

\def\d{\delta}
\def\e{\epsilon}
\def\ve{\varepsilon}
\def\f{\phi}
\def\g{\gamma}
\def\k{\kappa}
\def\l{\lambda}
\def\r{\rho}
\def\s{\sigma}
\def\t{\tau}
\def\th{\theta}
\def\vt{\vartheta}
\def\vp{\varphi}
\def\z{\zeta}
\def\o{\omega}
\def\D{\Delta}
\def\L{\Lambda}
\def\G{\Gamma}
\def\O{\Omega}
\def\S{\Sigma}
\def\Th{\Theta}
\def\del #1{\frac{\partial^{#1}}{\partial\l^{#1}}}
\def\h{\eta}
\def\AA{{\cal A}}
\def\BB{{\cal B}}
\def\CC{{\cal C}}
\def\DD{{\cal D}}
\def\EE{{\cal E}}
\def\FF{{\cal F}}
\def\GG{{\cal G}}
\def\HH{{\cal H}}
\def\II{{\cal I}}
\def\JJ{{\cal J}}
\def\KK{{\cal K}}
\def\LL{{\cal L}}
\def\MM{{\cal M}}
\def\NN{{\cal N}}
\def\OO{{\cal O}}
\def\PP{{\cal P}}
\def\QQ{{\cal Q}}
\def\RR{{\cal R}}
\def\SS{{\cal S}}
\def\TT{{\cal T}}
\def\VV{{\cal V}}
\def\UU{{\cal U}}
\def\VV{{\cal V}}
\def\WW{{\cal W}}
\def\XX{{\cal X}}
\def\YY{{\cal Y}}
\def\ZZ{{\cal Z}}
\def\A{{\cal A}}

\def\ba{{\backslash}}
\def\cov{\hbox{\rm cov}}
\def\sfrac#1#2{{\textstyle{#1\over #2}}}
\def\rank{\hbox{\rm rank}}
\def\limlaw{\buildrel \DD\over\rightarrow}
\def\wh{\widehat}
\def\TH(#1){\label{#1}}\def\thv(#1){\ref{#1}}
\def\theo #1{\noindent{\bf Theorem {#1} }}
\def\remark{\noindent{\bf Remark: }}
\def\proposition #1{\noindent{\bf Proposition #1}}
\def\note#1{\footnote{#1}}

 \eqsec  
\title{Energy statistics in disordered systems:\\ The local
REM  conjecture and beyond %
\thanks{Research supported in part by the DFG
in the Dutch-German Bilateral Research Group ``Mathematics of Random Spatial
Models from Physics and Biology'' and by the European Science
Foundation in the Programme RDSES}%
}
\author{Anton Bovier
\address{
Weierstra\ss--Institut
f\"ur Angewandte Analysis und Stochastik\\
Mohrenstrasse 39\\
10117 Berlin, Germany\\
and\\
Institut f\"ur Mathematik\\
Technische Universit\"at Berlin\\
Strasse des 17. Juni 136\\
12623 Berlin, Germany\\
bovier@wias-berlin.de
}
\and
Irina Kurkova
\address{
Laboratoire de Probabilit\'es et Mod\`eles
Al\'eatoires\\
Universit\'e Paris 6\\
4, place Jussieu, B.C. 188\\
75252 Paris, Cedex 5, France\\
 kourkova@ccr.jussieu.fr
}
}
\maketitle
\begin{abstract}
Recently, Bauke and Mertens conjectured that
the local statistics of energies in random spin systems with discrete
spin space should in most circumstances be the same as in the random
energy model. Here we give necessary conditions for this hypothesis
to be true, which we show to hold in wide classes of examples: short
range spin glasses and mean field spin glasses of the SK type. We also
show that, under certain conditions, the conjecture holds even if
energy levels that grow moderately
 with the volume of the system are considered.
In the case of the Generalised Random energy model, we give a complete
analysis for the behaviour of the local energy statistics at all
energy scales. In particular, we show that, in this case, the REM
conjecture holds exactly up to energies $E_N<\b_c N$, where $\b_c$ is
the critical temperature. We also explain the more complex behaviour
that sets in at higher energies.

\end{abstract}
\section{Introduction}

In a recent paper \cite{BaMe}, Bauke and Mertens have formulated an
interesting conjecture on the behaviour of local energy level
statistics in disordered systems. Roughly speaking, their conjecture
can be formulated as follows. Consider a random Hamiltonian,
$H_N(\s)$, i.e.\ a random function from some discrete spin  space,
$\SS^{\L_N}$, of the form
\begin{equation}
H_N(\s) =\sum_{A\subset \L_N} \Phi_A(\s)
\label{Mer.1}
\end{equation}
 where $\L_N$ are finite subsets of $Z^{d}$ of cardinality, say,
$N$,
 the sum runs over some  set
  of finite subsets of $\L_N$
 and $\Phi_A$ are
random local interactions, typically of the form
\begin{equation}
\Phi_A(\s) =J_A\prod_{x\in A}\s_x
\label{Mer.2}
\end{equation}
where $J_A$, $A\subset \Z^d$, is a family of (typically
independent) random variables. In such a situation, for typical
$\s$, $H_N(\s) \sim \sqrt N$, while $\sup_{\s}H_N(\s)\sim N$.
Bauke and Mertens then ask the following question: Given a fixed
number, $E$, what are  the statistics of the values
$N^{-1/2}H_N(\s)$ that are closest to this number $E$, and how are
configurations, $\s$, for which these good approximants of $E$ are
realised, distributed on $\SS^{\L_N}$? Their conjectured answer is
rather simple: find $\delta_{N, E}$ such that $\P(
|N^{-1/2}H_N(\s)-E|\leq  b \delta_{N, E}) \sim |\SS|^{-N} b$; then
the collection of points $ \delta_{N, E}^{-1} |N^{-1/2}H_N(\s)-E|$
over
     all  $\s\in \SS^{\L_N}$
converges to a Poisson point process on $\R_+$.
   Furthermore,  for any finite
$k$, the $k$-tuple of configurations
   $\s^1,\s^2,\dots, \s^k$, where the $k$ best approximations
are realised, is such that all of its elements have maximal Hamming
distance between each other. In other words, the asymptotic behavior of these
best approximants of $E$ is the same, as if the random variables
$H_N(\s)$ were all independent Gaussian random
  variables with zero mean and variance $N$, i.e.\ as if we were dealing with the
random energy model (REM) \cite{Der1}, that is why Bauke and Mertens call
this ``universal REM like behaviour''.

A comparable result, which motivated this problem, had previously been
 found by Mertens
\cite{Mer1} (for rigorous proofs see \cite{BCP, BCMP}) in the particular case of the {\it number partitioning
  problem}, where
\begin{equation}
H_N(\s)= \sum_{i=1}^N X_i\s_i
\label{Mer.3}
\end{equation}
with $X_i$ i.i.d.\ random variables uniformly distributed on
$[0,1]$,
 $\s_i\in \{-1,1\}$;  one is interested in the distribution of
energies near the value zero, which corresponds to an optimal
partitioning of the $N$ random variables $X_i$ into two groups, such
that their sum in each group is as similar as possible (a
 generalisation to other values of $E$ was investigated in \cite{BCMN}).

In \cite{BK04} we generalised this result to the case of the
$k$-partitioning problem, where the random function to be considered
is actually vector-valued (consisting of the vector of differences
between the sums of the random variables in each of the $k$ subsets of
the partition). To be precise, we considered the special case where
the subsets of the partition are required to have the same
cardinality, $N/k$ (restricted $k$-partitioning problem).
The general approach to the proof we developed in that paper sets the
path towards the proof of the conjecture by Bauke and Mertens that we
will
present here.

The universality conjecture suggests that correlations are
irrelevant for the properties of the local energy statistics of
disordered systems for energies near ``typical energies''. On the
other hand, we know that correlations must play a role for the
extreme energies near the maximum of $H_N(\s)$. Thus there are two
questions beyond the original conjecture that naturally pose
themselves: (i) assume we consider instead of fixed $E$,
$N$-dependent energy levels, say, $E_N=N^{\a}C$. How fast can we
allow $E_N$ to grow for the REM-like behaviour to hold? and (ii)
what type of behaviour can we expect once $E_N$ grows faster than
this value?  We will see that the answer to the first question
depends on the properties of $H_N$, and we will give an answer in
a class of examples related to the Sherrington-Kirkpatrick (SK)
spin-glass models. The answer to question (ii)  requires a
detailed understanding of $H_N(\s)$ as a random process, and we
will be able to give a complete answer on only in the case of the
GREM, when $H_N$ is a hierarchically correlated Gaussian process.
In this note we explain why the local REM conjecture is true in a
wide class of models, to what extend it can be generalised. In the
case of the GREM, we give a complete analysis of the local energy
statistics at all energy scales. We show that the REM conjecture
holds for all energies $E_N<\b_c N$, where $\b_c$ is the critical
temperature. Beyond that, there appear more complicated limiting
processes (compound Poisson processes) that reflect the onset of
an intrinsic geometrical structure. We explain this phenomenon in
detail. The emergence of $\b_c$, the critical value for the REM
conjecture in the GREM, suggests that possibly $\b_c N$ is a
general threshold for the REM-like behaviour in other models as
well, in particular in the SK models. It will be a challenging
problem for future work to investigate this possibility.

The technical details of our proofs will be presented elsewhere. Here
we will concentrate on anon-technical exposition of the ideas behind
these proofs.

{\bf Acknowledgements:}  We would like to thank Stephan Mertens for interesting
 discussions.

\section{The local REM}

Our approach to the proof of the REM conjecture is based on
the following general fact about random variables.
  Let $V_{i,M} \geq 0$,
    $i\in \N$, be a family of positive  random variables with
    identical distributions, that are normalized, s.t.
\begin{equation}
  (V_{i_j, M} <b)\sim \frac bM.
\label{lett.1}
\end{equation}
In the case of independent random variables, it would follow, that
the number of $V_{i,M}$ that are smaller than $b$ will have
Binomial distribution  with parameters $M$, $b/M$.
 If (\ref{lett.1}) holds for all $b\in \R_+$, if we plot the set  of
all points $V_{i,M}$, the number of points within any subset, $A
\subset \R_+\ba\{0\}$ will converge to a random variable with
Poisson distribution, with parameter the volume of the set $A$.
Moreover, if $A$ and $B$ are two disjoint subsets, then the
numbers of points within each respective set will be independent.
This means that the random set of points $V_{i,M}$ converges to a
{\it Poisson point process} on $\R_+\ba\{0\}$. The question   is
under which conditions this is still true, if the random variables
$V_{i,M}$ are correlated. It turns out that a very useful
sufficient condition is that, for any fixed number, $\ell$,
 average over all choices of collections of $\ell$ different the variables
$V_{i_1,M},\dots ,V_{i_\ell,M}$, the probabilities that all $\ell$ are
below thresholds, $b_i$, behaves as in the independent case, when $M$
goes to infinity, i.e.
\begin{equation}
\lim_{M\uparrow\infty}\sum_{(i_1,\dots,i_\ell)\subset
\{1,\dots,M\} }\P
  (\forall_{j=1}^\ell\,
V_{i_j, M} <b_j)\to \prod\limits_{j=1}^{\ell}b_{j}.
\label{maincond}
\end{equation}
 where the sum is taken
   over all possible  sequences of {\sl different\/}
    indices
  $(i_1,\dots, i_\ell)$.

\medskip
\remark A proof of this result  can be found in Chapter
    13 of \cite{B}.
\medskip
Naturally, we would apply this theorem with $V_{i,M}$  given by
$|N^{-1/2}H_N(\s)-E_N|$, properly normalised.

It is useful to think of the random variables $Y_N(\s)\equiv
N^{-1/2}H_N(\s)$ as Gaussian  random variables with variance one.
This holds, if the couplings, $J_A$, are Gaussian.  Otherwise, it
is one of the main steps of the proof to show that they converge,
in a strong sense, to Gaussians. This  requires some continuity
assumptions on the distributions of the couplings, but then holds
in great generality, if $E_N=E$ is independent of $N$. If energies
$E_N $ that go to infinity are considered, this problem is much
harder.

Let us see what needs to be checked in order for the above results to apply.
Consider a product space $\SS^N$ where $\SS$ is a finite set.
 We define on $\SS^N$  a real-valued random   process, $Y_N(\s)$,
  $\s \in \SS^N$.
Assume for simplicity that
 $
\E Y_N(\s)=0,\ \    \E(Y_N(\s))^2=1.
$
Define on $\SS^N$
\begin{equation}
b_N(\s,\s')\equiv \cov(Y_N(\s),Y_N(\s')).
\label{ab.abs.1}
\end{equation}

One problem we have to deal with from the outset are symmetries.
Let $G$ be the group of automorphisms on $\SS_N$, such that, for
$g\in G$, $Y_N(g\s)=Y_N(\s)$.  Then it is clear that we should only
consider the      residual classes of $\SS^N$ modulo this
group\note{In the special case $E=0$, we need even consider a larger
group that leaves the modulus of $Y_N$ invariant.}, which
we denote by $\S_N$.
Let us consider energies
\begin{equation}
E_N= c N^{\a},\ \ \ c,\a \in \R,\ \  0\leq \a<1/2,
\label{ab.abs.2}
\end{equation}
and  define the sequence
 \begin{equation}
\d_N= \sqrt{\sfrac {\pi}2} e^{ E_N^2/2} |\S_N|^{-1}.
\label{ab.abs.2.1}
\end{equation}
   Note that $\d_N$ is   exponentially small in  $N \uparrow \infty$,
   since $\a<1/2$. This sequence
   is chosen such that for any $b\geq 0$,
\begin{equation}
\lim_{N\uparrow\infty}|\S_N| \P(|Z_N(\s)-E_N|< b \d_N ) =b.
\label{ab.abs.2000}
\end{equation}

  For $\ell\in\N$ and any collection, $\s^1,\dots,\s^\ell\in
  \S_N^{\otimes \ell}$,
we denote by  \\  \ $B_N(\s^1,\dots,\s^\ell)$ \ \ the covariance matrix
with elements
\begin{equation}
b_{i,j}(\s^1,\dots, \s^{\ell})\equiv b_N(\s^i,\s^j).
\label{ab.abs.41}
\end{equation}

\begin{itemize}
\item[(i)]
Let $\RR_{N,\ell}^\eta$ denote the set where all covariances are
small, i.e. with $\eta>0$
\begin{equation}
\RR^\eta_{N,\ell}\equiv
 \{ (\s^1,\dots,\s^\ell)\in\S_N^{\otimes \ell}: \, \forall_{1\leq
i<j\leq \ell}\ |b_N(\s^i,\s^j)|\leq N^{-\eta} \}.
\label{ab.abs.6}
\end{equation}
This set will in any reasonable model exhaust almost the entire
configuration space. We will assume that
there exists a continuous decreasing
   function, $\rho(\eta)>0$,
 on $]\eta_0, \tilde \eta_0[$  (for some $
   \tilde \eta_0 \geq \eta_0>0$),  and a function,
    $\mu(\eta)>0$, on $]\eta_0, \tilde \eta_0[$,  such that
\begin{equation}
|\RR^\eta_{N,\ell}|\geq \left(1-\exp\left({- \mu(\eta) N^{\rho(\eta)}}
\right)\right) |\S_N|^\ell.
\label{ab.abs.4}
\end{equation}

\item[(ii)] Our main worry would then be degenerate situations where
the covariance matrix is singular. We need that the number of such
configurations is very small, in the following sense.
For $\ell \geq 2$,  $r=1,\dots, \ell-1$,
set
\begin{eqnarray}\nonumber
 \LL^\ell_{N,r}= \Big\{(\s^1,\dots,\s^\ell)\in
   \S_N^{\otimes \ell}& : & \  \forall_{1\leq i<j \leq \ell}
      |Y_N(\s^i)| \ne |Y_N(\s^j)|,  \\ \nonumber
&&\ \ \ \  \rank(B_N(\s^1,\dots, \s^{\ell}))=r
 \Big\}
\end{eqnarray}
Then there exists $d_{r, \ell}>0$, such that,
  for all $N$ large enough,
\begin{equation}
|\LL^\ell_{N,r}|\leq |\S_N|^r e^{-d_{r,\ell} N }.
\label{ab.abs.8}
\end{equation}
This is the most critical assumption that is not always easy to verify.
\item[(iii)] Finally, we must ensure that even in the degenerate
cases, the probabilities considered are under control. We ask that
for  $\ell\geq 1$, any  $r=1,2,\dots, \ell$,
    and any $b_1,\dots, b_{\ell}\geq 0$, there exist  constants,
$p_{r, \ell}\geq 0$  and
    $Q< 0$, such that,
 for any $\s^1,\dots, \s^{\ell} \in \S_N^{\otimes \ell}$,
such that  \\
    $\rank(B_N(\s^1,\dots,\s^\ell)) =r$,
\begin{equation}
\P \big (\forall_{i=1}^{\ell} :
   |Y_N(\s^i)-E_N |<\d_N b_i \big)
      \leq   Q \d_N^r  N^{p_{r, \ell}}.
\label{ab.abs.7}
\end{equation}
\end{itemize}

\theo{\TH (ABS.1)}  {\it  Under the assumptions above,
  if  $\a \in [0, 1/2[$  is such that,
    for some $\eta_1\leq\eta_2 \in ]\eta_0, \tilde \eta_0[$,
\begin{equation}
\a <\eta_2/2,
\label{ab.abs.9}
\end{equation}
\begin{equation}
\a <\eta/2+\rho(\eta)/2,  \,  \forall \eta \in ]\eta_1 , \eta_2[,
\label{ab.abs.10}
\end{equation}
and
\begin{equation}
\a<\rho(\eta_1)/2.
\label{ab.abs.10a}
\end{equation}
       Furthermore,  assume that, for any  $\ell \geq 1$,
 any $b_1,\dots, b_{\ell}>0$,  and\\
    $(\s^1,\dots, \s^{\ell}) \in \RR^{\eta_1}_{N, \ell}$,
\begin{eqnarray}
&\P& \left(\forall_{i=1}^{ \ell}:\
    |Y_N(\s^i)-E_N |< \d_N b_i \right)\\\nonumber
=
&\P& \left(\forall_{i=1}^{ \ell }:\
   |Z_N(\s^i)-E_N |<  \d_N b_i \right)+o(|\Sigma_N|^{-\ell}).
\label{ab.abs.5}
\end{eqnarray}
Then,  the point process,
\begin{equation}
P_N\equiv \sum_{\s\in \S_N} \d_{\{ \delta_N^{-1} |Y_N(\s)-E_N|\}}
\rightarrow\PP
\label{ab.abs.11}
\end{equation}
  converges weakly to the standard Poisson point process
 $\PP$ on $\R_+$.

\noindent Moreover, for any $\e>0$ and any $b\in \R_+$,
the probability that there exists two configuarations, $\s,\s'$, such
that
$|b_N(\s,\s')|>\e$ and  $ | Y_N(\s)-E_N|\leq |Y_N(\s')-E_N|\leq \d_Nb$,
tends to zero, as $N\uparrow\infty$.
}

Theorem \thv(ABS.1)  is proven by verifying that the conditions
(\ref{maincond}) are verified
for $V_{i,M}$ given by $\delta_N^{-1}|Y_{N}(\s)-E_N|$, i.e. that
\begin{equation}
\sum_{(\s^1,\dots,\s^\ell) \in \S_N^{\otimes l}}
\P \left(\forall_{i=1}^{\ell}:
 |Y_N(\s^i)-E_N|< b_i \d_N \right) \rightarrow b_1\cdots b_\ell.
\label{ab.abs.13}
\end{equation}
The formal proof is given in \cite{BK05a}.

The second assertion of the theorem is elementary:
 the sum of terms $\P(\forall_{i=1}^2 : |Y_N(\s^i)-E_N| <\delta_N b)$
  over all pairs,
    $(\s^1, \s^2) \in \Sigma_{N}^{\otimes 2}
  \setminus \RR_{N, 2}^{\eta_1}$, such that  $\s^1 \ne \s^2$,
 converges to zero exponentially fast.

%

  Finally, we  remark that the
  results  can be extended
to the case when  $\E Y_N(\s)\ne 0$, if $\a=0$, i.e. $E_N=c$.
Note that, e.g., the unrestricted number partitioning problem falls
  into this class.
 To see this, let $Z_N(\s)$ be the Gaussian process
  with the same mean and covariances as $Y_N(\s)$.
Let us consider both the covariance matrix, $B_N$, and the mean of
  $Y_N$,
 $\E Y_N(\s)$, as random variables on the probability space
$(\S_N,\BB_N,\E_\s)$, where $\E_\s$ is the uniform law
on $\S_N$.
 Assume that, for any $\ell\geq 1$,
\begin{equation}
B_N(\s^1,\dots, \s^{\ell}) \limlaw I_d, \ \ N \uparrow \infty,
   \label{ab.z1}
\end{equation}
   where $I_d$ denotes the identity matrix,  and
\begin{equation}
\E Y_N(\s) \limlaw D, \ \ N \uparrow \infty,
 \label{ab.z2}
\end{equation}
 where $D$ is some random variable $D$.
Let
\begin{equation} \widetilde \delta_N =\sqrt{\sfrac \pi2}K^{-1}
    |\Sigma_N|^{-1}.
\label{vbvb}
\end{equation}
 where
\begin{equation}
K\equiv \E e^{-(c-D)^2/2}.
\label{ab.z3}
\end{equation}

\theo{\TH(ABS.3)}{\it  Assume that, for some
  $R>0$, $|\E Y_N(\s)| \leq N^R$, for all $\s \in \S_N$.
Assume that (\ref{ab.abs.4})  holds for some $\eta>0$ and
  that (ii) and (iii) of Assumption~A are valid.
Assume that there exists a set, $\QQ_{N} \subset \RR^{\eta}_{N,
\ell}$,
 such that
 (\ref{ab.abs.5}) is valid   for any
   $(\s^1, \dots, \s^{\ell})\in \QQ_{N}$, and that
$|\RR_{N,\ell}^{\eta} \setminus \QQ_{N}|
   \leq |\Sigma_N|^{\ell} e^{-N^{\gamma}}$,
   with some $\gamma>0$.
 Then, the point process
\begin{equation}
P_N\equiv \sum_{\s\in \S_N} \d_{ \widetilde
   \delta_N^{-1} |Y_N(\s)-E_N|}\rightarrow\PP
\label{ab.abs.111}
\end{equation}
    converges weakly to the standard Poisson point process
 $\PP$ on $\R_+$ . }

\section{Examples}

The assumptions of our theorem are verified in a
wide class of physically relevant models. The examples we verified
explicitly in \cite{BK05a} are:  1) the Gaussian $p$-spin SK
models, 2) SK-models with non-Gaussian couplings,
and 3) short-range spin-glasses. In the last two examples we consider
only  the case $\a=0$.

\subsection{$p$-spin Sherrington-Kirkpatrick models,
    $0\leq \a<1/2$}

 In this subsection we illustrate  our general theorem in
 the class of Sherrington-Kirkpatrick models.
  Consider $\SS=\{-1, 1\}$.
\begin{equation}
  H_N(\s)=\frac{\sqrt{N}}{ \sqrt{ N \choose p }}\sum_{1\leq
i_1<i_2<\dots <i_p
 \leq N }
   J_{i_1,\dots, i_p} \s_{i_1}\s_{i_2}\cdots \s_{i_p}
\label{ab.tt1.3}
\end{equation}
  is the Hamiltonian of the $p$-spin Sherrington-Kirkpatrick model,
 where \\
$J_{i_1,\dots, i_p}$ are independent
    standard Gaussian random variables.

 The following elementary proposition concerns the symmetries to
the Hamiltonian.

\proposition{\TH(pr)}{\it
  Assume that, for any $0<i_1<\dots <i_p\leq N$,
     $\s_{i_1}\cdots \s_{i_p} =\s'_{i_1}\cdots \s'_{i_p}$.
  Then, if $p$ is pair,
   either $\s_i=\s_i'$, for all $i=1,\dots, N$, or $\s_i=-\s'_i$,
   for all $i=1,\dots, N$, and, if $p$ is odd, then
     $\s_i=\s_i'$, for all $i=1,\dots, N$.
   Assume that, for any $0<i_1<\dots <i_p\leq N$,
      $\s_{i_1}\cdots \s_{i_p} = -\s'_{i_1}\cdots \s'_{i_p}$.
 This is impossible, if $p$ is pair  and
    $\s_i=-\s_i'$, for all $i=1,\dots, N$, if
    $p$ is odd. }

  This proposition allows us to construct
   the space $\Sigma_N$:
 If $p$ is odd and $c \ne 0$,
  $\Sigma_N=\SS^N$, thus  $|\Sigma_N|=2^N$.
         If $p$ is even,  or $c=0$,
      $\Sigma_N$ consists of equivalence classes
   where configurations
    $\s$ and $-\s$ are identified, thus
      $|\Sigma_N|=2^{N-1}$.

\theo{\TH(unp-not0)}{\it  Let $p\geq 1$ be odd.
    Let $\Sigma_N =\SS^N$.
          If $p=1$ and
   $\a \in [0, 1/4[$, and, if $p=3,5,\dots,$,
   and  $\a \in [0, 1/2[$, for any constant $c \in \R\setminus\{0\}$
   the point process
\begin{equation}
\PP_N\equiv
\sum_{\s \in \Sigma_N } \delta_{\{ \d_N^{-1}
   |N^{-1/2} H_N(\s) - c N^{\a } | \}  }
\label{pp1}
  \end{equation}
where
$\d_N=2^{-N} e^{+c^{2} N^{2\a }/2}
 \sqrt{\sfrac\pi 2}$,
converges weakly to the standard Poisson point process, $\PP$,  on $\R_+$.

Let $p$ be even. Let $\Sigma_N$ be the space of equivalence classes
   of  $\SS^N$ where $\s$ and $-\s$ are identified.
        For any
      $\a \in [0, 1/2[$ and any constant, $c\in \R$,
   the point process
\begin{equation}
\PP_N\equiv\sum_{\s \in \Sigma_N } \delta_{\{(2\d_N)^{-1}| N^{-1/2}
   H_N(\s)- c N^{\a} | \} }
\label{pp2n}
\end{equation}
 converges weakly to the standard Poisson point process, $\PP$, on $\R_+$.
  The result (\ref{pp2n}) holds true as well  in the
  case of $c=0$, for  $p$ odd.
}
\subsection{Short range spin glasses}

As a final example, we consider  short-range spin glass  models.
To avoid unnecessary
complications, we will look at models on the $d$-dimensional  torus, $\L_N$,
of length $N$. We consider Hamiltonians of the form
\begin{equation}
H_N(\s)\equiv - N^{-d/2}\sum_{A\subset\L_N} r_A J_A \s_A
\label{srsg.1}
\end{equation}
where e $\s_A\equiv \prod_{x\in A}\s_x$, $r_A$ are given constants,
and $J_A$ are random variables.
We will introduce some notation:

\begin{itemize}
\item[(a)] Let $\AA_N$ denote the set of all $A\subset \L_N$, such
that $r_A\neq 0$.
\item[(b)] For any two subsets, $A,B\subset\L_N$, we say that
$A\sim B$, iff there exists $x\in\L_N$ such that $B=A+x$. We denote by
$\AA$ the set of equivalence classes of $\AA_N$ under this relation.

We will assume that the constants, $r_A$, and the random variables, $J_A$,
satisfy the following conditions:
\item[(i)] $r_A=r_{A+x}$, for any $x\in \L_N$;
\item[(ii)] there exists $k\in \N$,
such that any equivalence class in $\AA$ has a representative $A\subset\L_k$;
we will identify the set $\AA$ with a uniquely chosen set of
representatives contained in $\L_k$.
\item[(iii)] $\sum_{A\subset \L_N :} r_A^2 =N^d$.
\item[(iv)]
$J_A$, $A\in \Z^d$, are a family of independent
 random variables, such that
\item[(v)] $J_A$ and $J_{A+x}$
are identically distributed for any $x\in \Z^d$;
\item[(vi)] $\E J_A=0$ and $\E J_A^2=1$, and $\E J_A^3<\infty$;
\item[(vii)] For any $A\in\AA$, the Fourier transform
$\phi_A(s)\equiv\E \exp \left(i s J_A \right)$, of $J_A$ satisfies
 $|\phi_A(s)|=O(|s|^{-1})$ as $|s|\to  \infty$.
\end{itemize}

Observe that $\E H_N(\s)=0$,
\begin{equation}
b(\s,\s')\equiv N^{-d}\E H_N(\s)H_N(\s') =N^{-d}\sum_{A\in \L_N} r_A^2 \s_A\s_A'\leq 1
\label{srsg.2}
\end{equation}
where equality holds, if $\s=\s'$.

     Note that $Y_N(\s)=Y_N(\s')$
  (resp.  $Y_N(\s)=-Y_N(\s')$ ),
 if and only if, for all $A\in \AA_N$,
 $\s_A=\s_A'$ (resp. $\s_A=-\s_A'$).
E.g., in the standard Edwards-Anderson model,
with nearest neighbor pair interaction, if $\s_x$ differs from $\s_x'$
 on every second site,
$x$, then $Y_N(\s)=-Y_N(\s')$, and if $\s'=-\s$,  $Y_N(\s)=Y_N(\s')$.
In general, we will consider two configurations, $\s,\s'\in S^{\L_N}$,
 as equivalent, iff for all $A\in \AA_N$, $\s_A=\s_{A}'$. We denote the set
of these equivalence classes by $\S_N$. We will assume in the
sequel that there is a finite constant, $\Gamma\geq 1$, such that
$|\S_N|\geq 2^{N^d} \Gamma^{-1}$. In the special case of $c=0$,
the equivalence relation will be extended to include the case
$\s_A=-\s_{A}'$, for all $A\in \AA_N$. In most reasonable examples
(e.g. whenever nearest neighbor pair interactions are included in
the set $\AA$), the constant $\G\leq 2$ (resp.~$\G\leq 4$, if
$c=0$).

\theo{\TH(unppp)}{\it Let $c \in \R$, and  $\Sigma_N$  be
        the space of equivalence classes defined before.
Let $
\d_N\equiv  |\Sigma_N|^{-1}e^{c^2/2}\sqrt{\frac {\pi}2}$.
  Then the point process
\begin{equation}
\PP_N\equiv \sum_{\s \in \Sigma_N} \delta_{\{\d_N^{-1}
              |H_N(\s)-c|\} }
 \label{pll}
\end{equation}
  converges weakly to the standard Poisson point process on $\R_+$.

\noindent If, moreover, the random variables $J_A$ are Gaussian,
then, for any $c\in \R$, and $0\leq \a<1/4$, with  $ \d_N\equiv
|\Sigma_N|^{-1}e^{N^{2\a}c^2/2}\sqrt{\frac {\pi}2}$
 the point process
\begin{equation}
\PP_N\equiv \sum_{\s \in \Sigma_N} \delta_{\{\d_N^{-1}
              |H_N(\s)-cN^\a|\} }
 \label{pll.1}
\end{equation}
  converges weakly to the standard Poisson point process on $\R_+$.}

\section{Beyond REM behaviour}

    Let us briefly recall the definition of the GREM.
We consider parameters $\a_0=1<\a_1,\dots, \a_n<2$ with
  $\prod_{i=1}^n \a_i=2$,  $a_0=0<a_1,\dots, a_n<1$,
   $\sum_{i=1}^n a_i=1$. Let
     $X_{\s_1\cdots \s_l}$,
   $l=1,\dots, n$, be
  independent  standard Gaussian random variables indexed by
   $\s_1\dots\s_l\in \{-1,1\}^{N \ln(\a_1\cdots
    \a_l)/\ln 2}$.
  The Hamiltonian of the GREM is  $H_N(\s)\equiv\sqrt{N} X_\s$,
with
\begin{equation}
 X_{\s} \equiv\sqrt{a_1}X_{\s_1}+\cdots +\sqrt{a_n}X_{\s_1\cdots \s_n}.
\label{hmgrem}
\end{equation}
    Then $\hbox{cov}\, (X_{\s}, X_{\s'})=A(d_N(\s, \s'))$,
   where $d_N(\s, \s')=N^{-1}[\min \{i: \s_i\ne \s_i'\}-1]$,
  and $A(x)$ is a right-continuous
   step function  on $[0,1]$ with
    $A(x)=a_0+\cdots +a_i$, if
 $x\in [\ln(\a_0\a_1,\cdots\a_i)/\ln 2
    \,,\, \ln(\a_0\a_1,\cdots\a_{i+1})/\ln 2)$. We will assume here
   for simplicity that the linear envelope of the function $A$ is convex.

      To formulate our results, we also need
 to recall from \cite{BK1} (Lemma 1.2)
  the point process of Poisson cascades $\PP^l$
 on $\R^l$.  It is   best understood in terms of the
  following iterative construction. If $l=1$,  $\PP^1$
      is the Poisson point process on $\R^1$ with
 the intensity measure $e^{-x}dx$.
 To construct $\PP^l$, we place the process
  $\PP^{l-1}$ on the plane  of the first $l-1$ coordinates and through
    each of its points draw a straight line orthogonal to this plane.
         Then we put on each
  of these lines independently a Poisson point process with intensity
 measure $e^{-x}dx$. These points on $\R^l$
 form the process $\PP^l$.

Let us define the constants $d_l$, $l=0,1,\ldots,n$, where
$d_0=0$ and
\begin{equation}
d_l\equiv \sum_{i=1}^l\sqrt{ a_i 2\ln  \a_i}.
\label{ab.1}
\end{equation}
Finally, set, for $l=0,\dots,k-1$, as
\begin{equation}
D_l\equiv d_l+\sqrt {\frac {2\ln \a_{l+1}}{
a_{l+1}}}
\sum_{j=l+1}^k a_j.
\label{ab.mwgrem.2}
\end{equation}
    It is not difficult to verify that $D_0< D_1 <
         \cdots < D_{n-1}$.
Interestingly, the border of $D_0$ is the point $\b_c$, that
   is the critical
temperature of the  respective model.
     We are now ready to formulate the main result.

\theo {\it
If $|c| <D_0=\b_c$,
           then, the point process
\begin{equation}
\MM_N^0=\sum_{\s \in \Sigma_N} \delta_{\big\{2^{N+1}
(2\pi)^{-1/2}e^{-c_N^2 N/2}
        \big|X_{\s} - c_N \sqrt{N }\big|\big\} }
\label{m0}
\end{equation}
 converges to the Poisson point process with  intensity measure
  the Lebesgue measure.}

\theo{\it
 If for $l=1,\dots, n-1$, $D_{l-1}\leq c< D_l$, set
\begin{equation}
\tilde c_l =|c|- d_l,
\label{ccl}
\end{equation}
\begin{equation}
\beta_l= \frac{\tilde c_l}{a_{l+1}+\cdots + a_n}, \ \ \ \
 \ \gamma_i=  \sqrt{ a_i/ (2\ln  \a_i) }, \ \ i=1,\dots,l,
\label{not}
\end{equation}
and
\begin{equation}
   R_l(N)  =  \frac{2 (\a_{l+1}\cdots \a_k)^N
          \exp(-N\tilde c_l \beta_l/2) }{
       \sqrt{2\pi( a_{l+1}+\cdots + a_k)} }
      \prod\limits_{j=1}^l
 (4 N \pi \ln  \a_j)^{
       - \beta_l \gamma_j/2 }.
\label{ml}
\end{equation}
  Then, the point process
\begin{equation}
\MM_N^l = \sum_{\s \in \Sigma_N} \delta_{ \big\{ R_l(N)
       \big|\sqrt{a_1}X_{\s_1}+\cdots +\sqrt{a_n}X_{\s_1\dots \s_n}
    -  c\sqrt{N}\big| \big\} }
\label{pps}
\end{equation}
  converges to  mixed Poisson point process on $[0, \infty[$:
       given a realization of the random variable $\Lambda_l$,
 its intensity measure is $\Lambda_l dx$.
   The random variables $\Lambda_l$  is defined in terms of the Poisson
       cascades $\PP_l$ via
\begin{equation}
\Lambda_l= \int\limits_{\R^l}
      e^{\beta_l(\gamma_1 x_1+\cdots \gamma_l x_l)}\PP^l(dx_1,\dots, dx_l).
 \label{yyi}
 \end{equation}
}
The proof of this theorem is given in \cite{BK05b}. Here we  give a heuristic
interpretation of the main result.

Let us first look at (\ref{m0}). This statement corresponds to the
REM-conjecture of Bauke and Mertens \cite{BaMe}.
It is quite remarkable
        that this conjecture holds in the case of the GREM
   for energies of the form  $c N $
   (namely for $c \in \DD_0$).

In the REM \cite{Der1}, $X_{\s}$ are $2^N$ {\it independent\/}
  standard Gaussian random variables and
 a
statement (\ref{m0})
   would hold for all $c$ with  $|c|< \sqrt{2\ln 2}$:
 it  is a well
known result from the theory of
independent random variables \cite{LLR}. The value
$c=\sqrt{2\ln 2}$ corresponds to
 the maximum of $2^N$ independent standard Gaussian
random variables, i.e., $\max_{\s \in \Sigma_N } N^{-1/2} X_{\s}
  \to \sqrt{2\ln 2}$ a.s.
Therefore, at the level $c=\sqrt{2\ln 2}$,
   one has the emergence of
the extremal process. More precisely, the point process
\begin{equation}
\sum_{\s \in \S_N} \d_{\big\{\sqrt{2 N \ln 2 } \big(X_\s- \sqrt{2 N \ln 2 }+
    \ln (4\pi N \ln 2)/\sqrt{8 N \ln 2 } \big) \big\}},
\label{ab.grem.void}
\end{equation}
that is commonly written as
$\sum_{\s\in \S_N}\d_{u_N^{-1}(X_\s)}$
with
\begin{equation}
u_N(x)= \sqrt {2N\ln 2} -\frac {\ln ( 4\pi N \ln 2) }{2\sqrt{2N\ln
2}}+\frac{x}{\sqrt{2N\ln 2}},
\label{ik.1}
\end{equation}
converges to the Poisson point process $\PP^1$ defined above
(see e.g. \cite{LLR}).  For $c>\sqrt {2\ln2}$,
the probability that one of the $X_\s$ will  be outside
of  the domain $\{|x|< c\sqrt{N}\}$, goes to
zero, and thus it makes no sense to consider such levels.

In the GREM,  $N^{-1/2}\max_{\s \in \Sigma_N }X_{\s}$
  converges to the value
     $d_k \in \partial D_{k-1}$ (\ref{ab.1})
(see Theorem 1.5 of \cite{BK1}) that is generally smaller
   than $\sqrt{2\ln 2}$.
Thus  it makes no sense to consider levels
 with $c \not\in \overline D_{k-1}$.
However, the REM-conjecture is not true for all levels
  in $\DD_{k-1}$,
but only in the smaller
 domain $\DD_0$.

To understand the statement of the theorem outside $\DD_0$,
we need to recall how the extremal process in the GREM is
related to the Poisson cascades introduced above. Let us set
$\S_{Nw_l}\equiv \{-1,1\}^{Nw_l}$ where
\begin{equation}
w_l=\ln( \a_1\cdots  \a_l)/\ln 2
\label{www}
\end{equation}
 and define the functions
\begin{equation}
U_{l,N}(x)\equiv N^{1/2} d_l- N^{-1/2} \sum_{i=1}^l \g_i\ln( 4\pi N  \ln
\a_i )/2+
N^{-1/2}x
\label{a.grem.2}
\end{equation}
Set
\begin{equation}
\wh X^j_\s\equiv \sum_{i=1}^j\sqrt a_i X_{\s_1\dots\s_i}, \ \ \ \
\check X^j_\s
\equiv  \sum_{i=j+1}^{n}\sqrt a_i X_{\s_1\dots\s_i}.
\label{ab.grem.3.1}
\end{equation}
  From what was shown in \cite{BK1}, for any $l=1,\dots,n$,
    the point process,
\begin{equation}
\EE_{l,N}\equiv
\sum_{\hat \s \in \S_{Nw_l}}
 \delta_{U^{-1}_{l,N}(\wh X^{J_l}_{\hat \s} )}
\label{ab.grem.4}
\end{equation}
converges in law to the Poisson cluster process, $\EE_l$,  given in terms
of the Poisson cascade, $\PP^l$, as
\begin{equation}
\EE_l\equiv \int\limits_{\R^l} \PP^{(l)}(dx_1,\dots,dx_l)
\d_{\sum_{i=1}^l \g_ix_i}.
\label{ab.grem.5}
\end{equation}
In view of this observation, we can re-write the definition of the
process  $\MM_{N}^l$ as follows:

\begin{equation}
\MM_N^l = \sum_{\hat\s \in \Sigma_{w_lN}}
\sum_{\check\s\in \S_{(1-w_l)N}}
 \delta_{\big\{ R_l(N)
       \big|\check X^{J_l}_{\hat\s\check\s} -
\sqrt{N}  \big[|c|-d_l
-N^{-1}(\G_{l, N}-U^{-1}_{l,N}(\wh X^{J_l}_{\hat \s}))
\big] \big| \big\} },
\label{ab.grem.6}
\end{equation}
with the abbreviation
\begin{equation}
\G_{l, N}\equiv \sum_{i=1}^l\g_i \ln
  (4\pi N\ln\a_i )/2 \label{gl}
\end{equation}
  ($c$ is replaced by $|c|$
  due to the symmetry of the standard Gaussian
  distribution).
The normalizing constant, $R_l(N)$, is chosen such that, for any
finite value, $U$, the point process
\begin{equation}
\sum_{\check\s\in \S_{(1-w_l)N}}
 \delta_{\big\{ R_l(N)
       \big|\check X^{J_l}_{\hat\s\check\s} - \sqrt{N}  \big[|c|-d_l
-N^{-1}(\G_{l, N}-U) \big] \big| \big\},
}
\label{ab.grem.7}
\end{equation}
converges to the Poisson point processes on $\R_+$, with intensity
measure given by  $e^{U}$ times Lebesgue measure,
which is possible  because $c \in \DD_l \setminus
\overline{\DD_{l-1}}$,
  that is  $|c|-d_l$ is smaller that the limit of
   $N^{-1/2} \max_{\check \s\in \S_{(1-w_l)N}   }
   \check X^{J_l}_{\hat\s\check\s}$.
 This is
completely analogous to the analysis in the domain $\DD_0$.
Thus each term in the
sum over $\hat\s$ in (\ref{ab.grem.6}) that gives rise to a ``finite''
 $U^{-1}_{l,N}(\wh X^l_{\hat \s})$, i.e., to an element of the extremal
 process of $\wh X^l_{\hat \s}$, gives rise to one Poisson process
 with
a random intensity measure in the limit of $\MM_{N}^l$. This explains
how the statement of the theorem can be understood,
and also  shows what the geometry of the configurations
realizing these mixed Poisson point processes will be.

  Let us  add that,
        if $c \in \partial \DD_{k-1}$,  i.e.
$|c|=d_{k}$, then one has the emergence
  of the extremal point process (\ref{ab.grem.4}) with $l=k$, i.e.
\begin{equation}\label{ab.ende}
\sum_{\s \in \Sigma_N}
\delta_{\{\sqrt{N}(X_\s-d_k\sqrt{N}+N^{-1/2}\Gamma_{k, N}
)\} }\rightarrow\EE_k,
\end{equation}
see \cite{BK1}.


\end{document}